\documentclass[fleqn,twoside]{article}                                         
\usepackage{espcrc2}
\usepackage{graphicx} 

\newcommand{\AmS}{{\protect\the\textfont2
  A\kern-.1667em\lower.5ex\hbox{M}\kern-.125emS}}

\title{Localization and pattern formation in Wigner representation via multiresolution} 
\author{A.~N. Fedorova and M.~G. Zeitlin
\address{IPME, RAS, V.O. Bolshoj pr., 61,
 199178, St.~Petersburg, Russia\\
E-mail: zeitlin@math.ipme.ru, 
http://www.ipme.ru/zeitlin.html, 
http://www.ipme.nw.ru/zeitlin.html
}}

\begin{document}                                               

%%%%%%%%%%%%%%%%%%%%%%%%%%%%%%%%%%%%%%%%%%%%%%%%%%%%%%%%%%%%%%%%%%%%%%%%%%%%%%%%%%%

\thispagestyle{empty}

\begin{center}
\begin{tabular}{p{130mm}}

\begin{center}
{\bf\Large LOCALIZATION AND PATTERN FORMATION}\\
\vspace{5mm}

{\bf\Large IN WIGNER REPRESENTATION VIA} \\
\vspace{5mm}

{\bf\Large MULTIRESOLUTION}\\

\vspace{1cm}

{\bf\Large Antonina N. Fedorova, Michael G. Zeitlin}\\

\vspace{1cm}

{\bf\it
IPME RAS, St.~Petersburg,
V.O. Bolshoj pr., 61, 199178, Russia}\\
{\bf\large\it e-mail: zeitlin@math.ipme.ru}\\
{\bf\large\it e-mail: anton@math.ipme.ru}\\
{\bf\large\it http://www.ipme.ru/zeitlin.html}\\
{\bf\large\it http://www.ipme.nw.ru/zeitlin.html}
\end{center}

\vspace{1cm}
\begin{center}
\begin{tabular}{p{100mm}}
We present an application of variational-wavelet analysis to                  
quasiclassical calculations of solutions of Wigner equations related to           
nonlinear (polynomial) dynamical problems. (Naive) deformation                     
quantization, multiresolution representations and 
variational approach are the key points. 
Numerical calculations
demonstrates pattern formation from localized eigenmodes and 
transition from chaotic to localized (waveleton) types of behaviour.
\end{tabular}
\end{center}
\vspace{40mm}

\begin{center}
{\large Presented at VIII International Workshop on}\\
{\large  Advanced Computing and Analysis Techniques in Physics Research,} \\
{\large Section III "Simulations and Computations in }\\
{\large Theoretical Physics and Phenomenology"}\\
{\large ACAT'2002, June 24-28, 2002, Moscow}
\end{center}
\end{tabular}
\end{center}
\newpage

%%%%%%%%%%%%%%%%%%%%%%%%%%%%%%%%%%%%%%%%%%%%%%%%%%%%%%%%%%%%%%%%%%%%%%%%%%%%%%%%%%

\begin{abstract} 
We present an application of variational-wavelet analysis to                  
quasiclassical calculations of solutions of Wigner equations related to           
nonlinear (polynomial) dynamical problems. (Naive) deformation                     
quantization, multiresolution representations and 
variational approach are the key points. 
Numerical calculations
demonstrates pattern formation from localized eigenmodes and 
transition from chaotic to localized (waveleton) types of behaviour.
\end{abstract}               

\maketitle 

In this paper we consider the  applications of  
nu\-me\-ri\-cal\--\-ana\-ly\-ti\-cal technique based on local nonlinear harmonic analysis
(wavelet analysis) to  quasiclassical
quantization of nonlinear (polynomial/rational) dynamical problems 
which appear in many areas of physics. 
Our starting point is the general point of view of deformation 
quantization approach at least on
naive Moyal level [1].
So, 
for quantum calculations we need to find
an associative (but non-commutative) star product $*$ on the space of formal power series in $\hbar$ with
coefficients in the space of smooth functions such that
$\quad
f * g =fg+\hbar\{f,g\}+\sum_{n\ge 2}\hbar^n B_n(f,g). 
$\\\noindent
In this paper we consider calculations of Wigner functions $W(p,q,t)$ (WF) as the solution
of Wig\-ner equation [1]
\begin{eqnarray}
i\hbar\frac{\partial}{\partial t}W = H * W - W * H
\end{eqnarray}
corresponding to polynomial Hamiltonians\\ $H(p,q,t)$.
In general, equation (1)
is nonlocal (pseudodifferential) for arbitrary Hamiltonians but in case of 
polynomial Ha\-mil\-to\-ni\-ans
we have only a finite number of terms in the corresponding series.
For example, in stationary case after Weyl-Wigner mapping we have the following 
equation on WF in c-numbers [1]:
\begin{eqnarray}
&&\Big( \frac{p^2}{2m}+\frac{\hbar}{2i}\frac{p}{m}\frac{\partial}{\partial q}-
 \frac{\hbar^2}{8m}\frac{\partial^2}{\partial q^2}\Big)W(p,q)+\\
&& U\Big(q-\frac{\hbar}{2i}\frac{\partial}{\partial p}\Big)W(p,q)=\epsilon W(p,q)\nonumber
\end{eqnarray}
and after expanding the potential $U$ into the Taylor 
series we have two real finite partial differential equations.
Our approach is based on extension of our variational-wavelet approach [2], [3].
Wavelet analysis is some set of mathematical methods, which gives us the possibility to
work with well-localized bases in functional spaces and gives maximum sparse
forms for the general type of operators (differential, integral, pseudodifferential) 
in such bases.
We decompose the solutions of (1), (2) according to underlying hidden scales [2], [3] as 
\begin{eqnarray}
&&W(t,q,p)=\displaystyle\bigoplus^\infty_{i=i_c}\delta^iW(t,q,p)
\end{eqnarray}
where value $i_c$ corresponds to the coarsest level of resolution
$c$ in the full multiresolution decomposition
$V_c\subset V_{c+1}\subset V_{c+2}\subset\dots$.
As a result the solution of any Wigner-like equations has the 
following mul\-ti\-sca\-le\-/mul\-ti\-re\-so\-lu\-ti\-on decomposition via 
nonlinear high\--lo\-ca\-li\-zed eigenmodes, 
which corresponds to the full multiresolution expansion in all underlying  
scales starting from coarsest one
(polynomial tensor bases are introduced in [3]; ${\bf x}=(x,y,p_x,p_y)$ e.g.):
\begin{eqnarray}\label{eq:z}
&&W(t,{\bf x})=\sum_{(i,j)\in Z^2}a_{ij}{\bf U}^i\otimes V^j(t,{\bf x}),\\
&&V^j(t)=V_N^{j,slow}(t)+\sum_{l\geq N}V^j_l(\omega_lt), \quad \omega_l\sim 2^l \nonumber\\
&&{\bf U}^i({\bf x})={\bf U}_M^{i,slow}({\bf x})+
\sum_{m\geq M}{\bf U}^i_m(k_m{\bf x}), \ k_m\sim 2^m\nonumber
\end{eqnarray}
\begin{figure}[htb]
\centering
\includegraphics*[width=50mm]{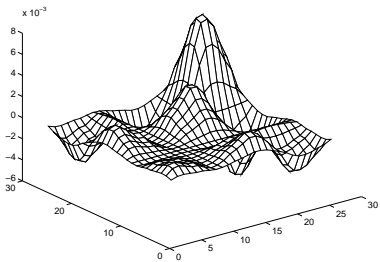}
\caption{Localized mode contribution to Wigner function.}
\end{figure}
\begin{figure}[htb]
\centering
\includegraphics*[width=50mm]{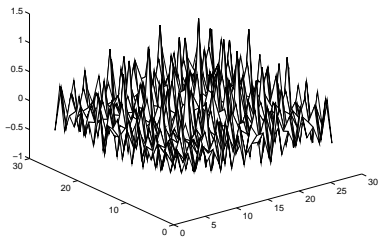}
\caption{Chaotic-like Wig\-ner function.}
\end{figure}
Formula (4) gives the expansion into the slow part
and fast oscillating parts for arbitrary N, M.  So, we may move
from coarse scales of resolution to the 
finest one for obtaining more detailed information about our dynamical process.
In contrast with different approaches formulae (3), (4) do not use perturbation
technique or linearization procedures.
So, by using wavelet bases with their best (phase) space/time      
localization properties we can describe localized (coherent) structures in      
quantum systems with complicated behaviour.
Modeling demonstrates the appearance of stable patterns formation from
high-localized coherent structures or chaotic behaviour.
Our (nonlinear) eigenmodes are more realistic for the modelling of 
nonlinear classical/quantum dynamical process  than the corresponding linear gaussian-like
coherent states. Here we mention only the best convergence properties of expansions 
based on wavelet packets, which  realize the minimal Shannon entropy property
and exponential control of convergence based on the norm introduced in [3].
Fig. 1 shows the high-localized eigenmode contribution to WF, while Fig. 2, 3 gives 
the representations for full solutions, constructed
from the first 6 eigenmodes and demonstrate stable localized 
pattern formation (waveleton) and chaotic-like behaviour outside of KAM-like region.
We can control the type of behaviour on the level of reduced algebraical variational 
system [3]. 
\begin{figure}[htb]
\centering
\includegraphics*[width=65mm]{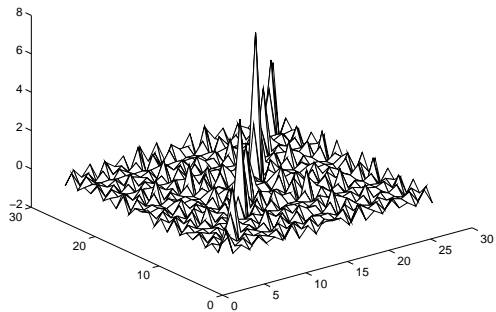}
\caption{Localized pattern-like (waveleton) Wig\-ner function.}
\end{figure}


\begin{thebibliography}{3}
                                                                                      
\bibitem{1}                                                                         
W. P. Schleich, Quantum Optics in Phase Space, Wiley, 2000

\bibitem{2}
A.N. Fedorova, M.G. Zeitlin,
Quantum Aspects of Beam Physics, 527-538, 539-550, World Scientific, 2002. 

\bibitem{3}
A.N. Fedorova, M.G. Zeitlin, arXiv preprints: physics/0206049, 0206050, 
0206051, 0206052, 0206053,
0206054, 0212066, nlin/0206024.
 

\end{thebibliography}
 \end{document}